\documentclass[letterpaper,prl,twocolumn,amsmath,amssymb,superscriptaddress,longbibliography]{revtex4-1}
\usepackage[utf8]{inputenc}
\usepackage[draft]{hyperref}
\usepackage{graphicx}% Include figure files
\usepackage{dcolumn}% Align table columns on decimal point
\usepackage{bm}
\usepackage{amssymb}
\usepackage{physics}
\usepackage{color}
\usepackage{soul}
\usepackage{epstopdf}
\usepackage{float}
\usepackage{array}
\begin{document}

\title{Light-induced rotation of dielectric microparticles around an optical nanofiber}

\author{Georgiy Tkachenko}
\email[Corresponding author: ]{georgiy.tkachenko@oist.jp}
\author{Ivan Toftul}
\altaffiliation[Also at ]{ITMO University, Birzhevaya liniya 14, 199034 St.-Petersburg, Russia}
\author{Cindy Esporlas}
\author{Aili Maimaiti}
\altaffiliation[Also at ]{Department of Physics, Chalmers University of Technology, 412 96 G\"oteborg, Sweden}
\author{Fam Le Kien}
\author{Viet Giang Truong}
\email[Corresponding author: ]{v.g.truong@oist.jp}
\author{S\'{i}le {Nic Chormaic}}
\altaffiliation[Also at ]{Institut N\'eel, Universit\'e Grenoble Alpes, F-38042 Grenoble, France}
\affiliation{Light-Matter Interactions Unit, Okinawa Institute of Science and Technology Graduate University, Onna, Okinawa 904-0495, Japan}

\date{\today}

\begin{abstract}
We experimentally demonstrate orbiting of isotropic, dielectric microparticles around an optical nanofiber that guides elliptically polarized fundamental modes. The driving transverse radiation force appears in the evanescent electromagnetic fields due to orbital angular momentum. The force direction is opposite to that of the energy flow circulation around the nanofiber. Our results verify the theoretically predicted negative optical torque on a sufficiently large particle in the vicinity of a nanofiber.
\end{abstract}

\vspace{2pc}
%\noindent{\it Keywords}: evanescent field, spin-dependent force, negative radiation force and torque, nanofiber

\maketitle
Spin angular momentum (SAM) carried by paraxial free-space beams of light can be transferred to a material object, causing it to rotate around its axis (i.e., spin), if the object is absorbing or anisotropic~\cite{friese_nature_1998}. In contrast, orbital angular momentum (OAM) in beams with optical vortices can even set isotropic, non-absorbing particles into rotation~\cite{oneil_prl_2002,garceschavez_prl_2003}. In nonparaxial light, SAM and OAM can couple, leading to, for example, orbiting of isotropic particles trapped by a tightly focused, nonvortex beam~\cite{zhao_prl_2007} and to observable, spin-dependent, transverse shifts of the light itself~\cite{baranova_jetp_1994,bliokh_prl_2008}. 
Symmetry breaking in a system consisting of a scattering object at the interface between two media, under oblique illumination, produces an interesting spin-dependent optomechanical effect~\cite{sukhov_natPhot_2015}.

Evanescent electromagnetic fields, which accompany total internal reflection and guiding of light, exhibit even more complicated spin-orbit interactions. In particular, aside from the common axial SAM associated with polarization, such fields exhibit a SAM component perpendicular to the wave vector~\cite{bliokh_nc_2014}. In addition, a material object in an evanescent field can experience a transverse spin-dependent force, as demonstrated experimentally by means of a nanocantilever~\cite{antognozzi_nPhys_2016} or an optically trapped Mie scattering particle~\cite{liu_prl_2018} placed near a total internal reflecting glass surface.

The evanescent field around an optical nanofiber~\cite{tong_oc_2012} guiding a quasi-circularly polarized fundamental mode is also expected to carry significant OAM that is transferable to material objects~\cite{le_kien_pra_2006}. In spite of numerous demonstrations of particle trapping, propulsion~\cite{brambilla_ol_2007,xu_njp_2012,maimaiti_sr_2015}, and binding~\cite{frawley_oe_2014,maimaiti_sr_2016} in the vicinity of optical nanofibers, exploration  of the rotational degree-of-freedom has never been reported in the literature. The main reason for this lack of experimental evidence was the uncertainty about the polarization of light at the waist of a nanofiber waveguide. This uncertainty has been lifted only recently~\cite{lei_prappl_2019,joos_oe_2019,tkachenko_arxiv_2019}. In this Letter, we present a clear demonstration of the spin-dependent optical torque by means of light-induced orbiting of isotropic microspheres around a single-mode optical nanofiber.

%------------------------------------------------
\begin{figure}
\centering
\includegraphics[width=1\linewidth]{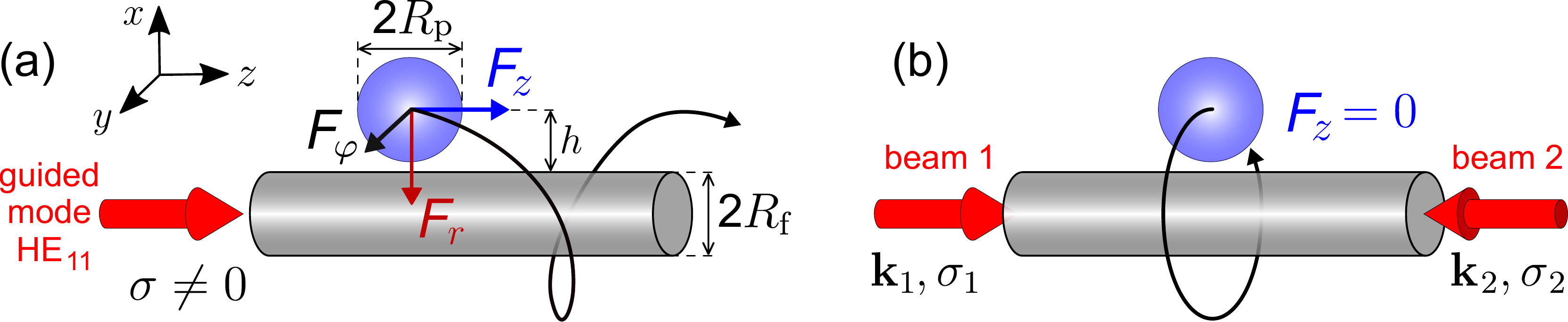}
\caption{(a)~An isotropic, dielectric particle in the evanescent field of an elliptically polarized, fundamental mode of an optical nanofiber. Due to the azimuthal optical force, $F_{\varphi}$, the particle can rotate around the fiber. (b)~We eliminate axial motion by using two counterpropagating beams with identical intensity profiles. The polarization in each beam is defined by $\sigma$, the projection of the SAM on the wave vector,~$\bf k$.}
\label{fig:concept}
\end{figure}
%------------------------------------------------

Let us consider the interaction between a spherical, dielectric particle (of radius $R_{\rm p}$) and the evanescent field of a single-mode optical nanofiber (of radius $R_{\rm f}$), as sketched in Fig.~\ref{fig:concept}(a). The electric part of an {\it elliptically} polarized guided mode is
\begin{equation}
    \bm{\mathcal{E}} = (\sqrt{1+\sigma}\,\bm{\mathcal{E}}_{p=+1}+{\mathrm e}^{i\phi}\sqrt{1-\sigma}\,\bm{\mathcal{E}}_{p=-1})/\sqrt{2}\,,
\end{equation}
where $\sigma\in[-1,1]$ is the ellipticity degree (i.e., the SAM projection on the wave vector, $\bf k$, in units of $\hbar$), $\phi\in[0,2\pi]$ determines the orientation of the symmetry axes of the polarization ellipse in the $xy$ plane, and $\bm{\mathcal{E}}_{p}=(e_r{\bf\hat{r}}+pe_{\varphi}{\hat{\bm{\varphi}}}+e_z{\bf\hat{z}}){\mathrm e}^{i\beta z+i p\varphi}$ is the electric part of the {\it quasicircularly} polarized guided mode with a polarization rotation index, $p=\pm1$~\cite{le_kien_pra_2013}. Here, $\beta$ is the propagation constant and $e_r$, $e_{\varphi}$, and $e_z$ are the cylindrical components of the mode-profile function of~$\bm{\mathcal{E}}_p$ with $p=+1$. The azimuthal component of the Poynting vector of the elliptically polarized guided mode is $S_{\varphi} = \sigma (e_z h_r^*-e_r h_z^*)/2$, where $h_r$ and $h_z$ are the components of the mode-profile function of the magnetic part, $\bm{\mathcal{H}}_{p}$, of the guided mode with a polarization index $p=+1$. Since the longitudinal field components, $e_z$ and $h_z$, are nonzero, we have $S_\varphi^{(p)}\equiv S_\varphi|_{\sigma=p}=p(e_zh_r^*-e_rh_z^*)/2\neq0$. We can show that $S_{\varphi}^{(p=+1)}>0$ and $S_{\varphi}^{(p=-1)}<0$ outside the nanofiber.

The light-induced force and torque on any object can be calculated if one knows the exact incident and scattered electromagnetic waves. In our problem, the incident wave (here, the evanescent field) is well-known~\cite{le_kien_pra_2017}. However, calculation of the scattered field on a Mie scattering particle is a challenging task. Following the generalized Lorenz-Mie theory, the incident field can be decomposed into vector spherical harmonics, and the scattered field is thus found by application of boundary conditions~\cite{barton_jap_1988,almaas_josab_1995}. Thence, the force and torque, respectively, can be found by integration of the linear and angular momenta over a surface enclosing the object.

The force exerted on a scattered particle near a nanofiber guiding a quasicircularly polarized fundamental mode can be decomposed into the axial ($F_z$), radial ($F_r$), and azimuthal ($F_\varphi$) components~\cite{le_kien_pra_2013}, see Fig.~\ref{fig:concept}(a).
Under $F_r$, the particle is attracted to the fiber surface and stays at $r=\sqrt{x^2+y^2}\geq(R_{\rm f}+R_{\rm p})$ (the inequality being due to surface roughness and Brownian motion). In this work we aim at detection of the azimuthal force, $F_\varphi$, which sets the particle into orbital motion around the fiber.
Since Brownian motion breaks mechanical contact between the particle and the fiber, the contribution from the intrinsic radiation torque to the azimuthal motion of the particle is expected to be negligible.
According to our calculations, $F_\varphi$ is much smaller than the axial force, $F_z$, which propels the particle towards $z>0$. In order to prevent $F_z$ from hindering detection of the light-induced rotation, we eliminate the axial motion by launching a second ${\rm HE}_{11}$ mode propagating towards $z<0$ into the nanofiber, with a power equal to that of the initial mode. This is realized experimentally by coupling non-interfering, Gaussian laser beams into the opposite pigtails of the tapered fiber, see Fig.~\ref{fig:concept}(b).

In principle, the rotation under $F_\varphi$ could be studied if beam~1 were elliptically polarized ($\sigma_1=\sigma\neq0$) and beam~2 were linearly polarized ($\sigma_2=0$). However, such a beam~2 would produce a mode with an axially asymmetric intensity profile~\cite{le_kien_oc_2004} and the particle would tend to stop at the `hot spots', unless $|\sigma_1|\approx1$. Since we consider the complete spectrum of $\sigma$, we set the polarization of beam~2 to also be elliptical, with $\sigma_2=-\sigma_1$. In this case, the total azimuthal force is the sum of the contributions from both beams.

Once $F_\varphi$ is known, the orbiting frequency of the particle at equilibrium can be easily calculated from the force balance equation, $F_\varphi + F_{\rm fr} = 0$, where $F_{\rm fr}$ is the friction. In our experiments, the particle is immersed in water, which produces a friction of $F_{\rm fr} = -\gamma v$, where $v$ is the linear velocity of the particle's center and $\gamma$ is the drag coefficient. If the particle were in a laminar flow far from any borders, the drag could be adequately described by the Stokes approximation: $\gamma = \gamma_{S} = 6\pi\eta R_{\rm p}$, where $\eta$ is the dynamic viscosity of the fluid ($\eta\approx1$~mPa~s for water at room temperature). However, the particle is essentially in contact with the fiber surface and this must be taken into consideration when estimating~$\gamma$. As demonstrated by Marchington~et~al.~\cite{marchington_oe_2008}, an appropriate description of the friction for a microsphere in the evanescent field (in that work, at the surface of a glass prism) can be obtained using the lubrication correction derived by Krishnan and Leighton~\cite{krishnan_pf_1995}:
\begin{equation}
    \gamma_{\rm KL}=-\gamma_{\rm S}\left[\frac{8}{15}{\rm ln}\left(\frac{h-R_{\rm p}}{R_{\rm p}}\right)-0.9588\right]\,,
    \label{eq:gamma_KL}
\end{equation}
where the distance $h=r-R_{\rm f}$ (see Fig.~\ref{fig:concept}(a)) depends on the particle surface roughness. We note that Eq.~\ref{eq:gamma_KL} is only valid for a limited size range, $R_{\rm p}>0.25\,\mu$m~\cite{krishnan_pf_1995}.
%For example, with $R_{\rm p}=2.5\,\mu$m one can assume $h-R_{\rm p}\approx10$~nm~\cite{marchington_oe_2008}.
The absolute value of the particle rotation frequency around a fiber when both beams are circularly polarized (CP) can thus be expressed as
\begin{equation}
    |f_{\rm CP}| = \frac{|v|}{2\pi (h+R_{\rm f})} = \frac{|F_{\varphi}|}{2\pi\gamma_{\rm KL}(h+R_{\rm f})}\,.
    \label{eq:f_CP}
\end{equation}
As follows from our simulations, in the general case of elliptical polarization (EP), the azimuthal force and the corresponding frequency, $f_{\rm EP}$, are proportional to $\sigma=\sigma_1$, with opposite signs:
\begin{equation}
    f_{\rm EP} = -\sigma|f_{\rm CP}|\,,
    \label{eq:f_EP}
\end{equation}
and this result is consistent with the theoretical findings of Le~Kien and Rauschenbeutel~\cite{le_kien_pra_2013}, for the relevant range of the size parameter, $n_{\rm m}kR_{\rm p}$, where $n_{\rm m}$ is the refractive index of the medium. For convenience, we normalize the rotation frequency by the total optical power $P$. That is, we use $\tilde{f}_{\rm CP,~EP}=f_{\rm CP,~EP}/P$.

%------------------------------------------------
\begin{figure}
\centering
\includegraphics[width=1\linewidth]{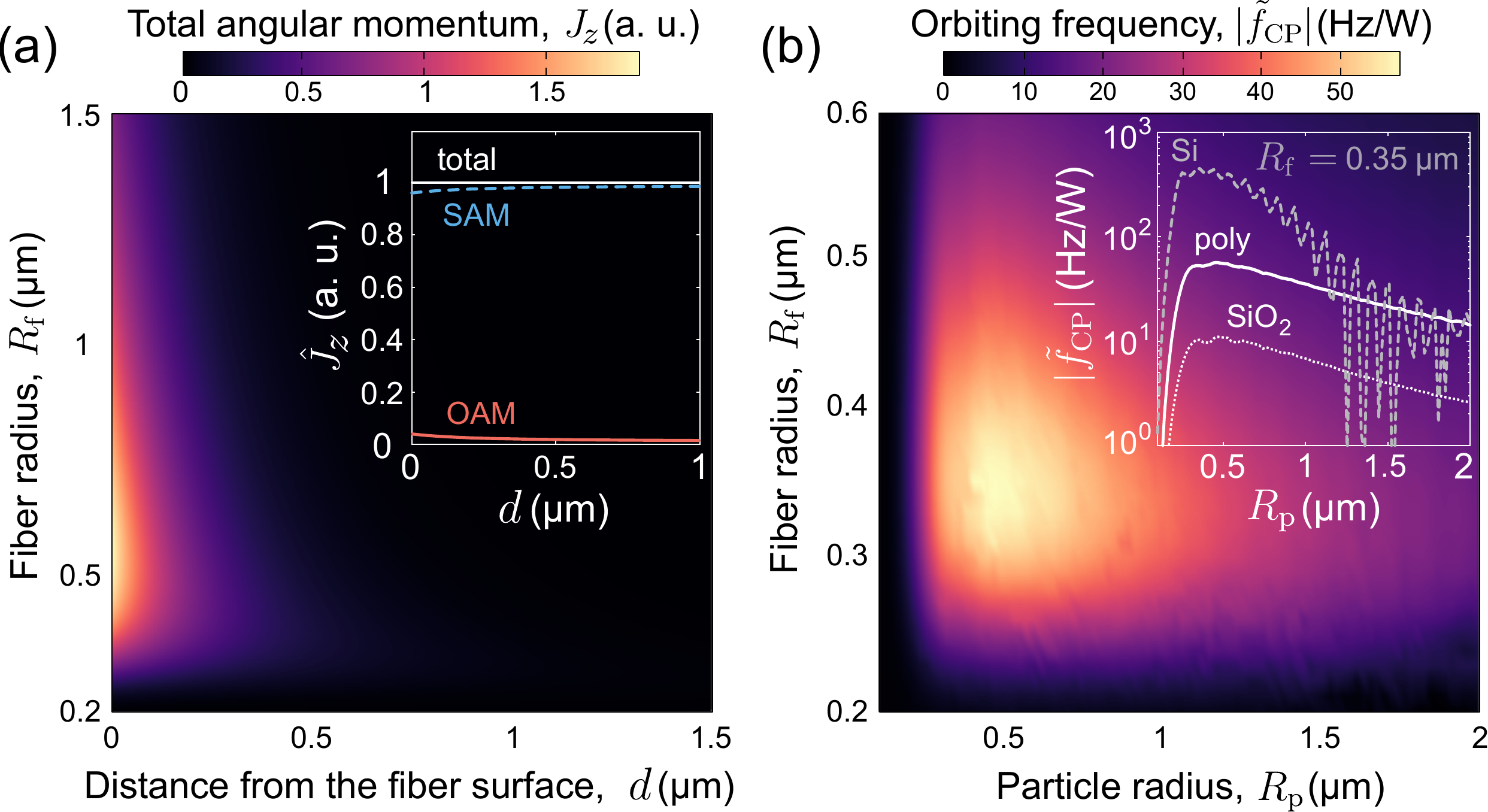}
\caption{Simulation results. (a)~Density of the total angular momentum of light near a nanofiber (in water) guiding a fundamental mode  with $\sigma=1$. Inset: total angular momentum per photon and its orbital and spin components. (b)~The orbiting frequency for a polystyrene particle, as a function of radii of the particle and the fiber. Inset: frequency at the optimum fiber radius ($R_{\rm f}=0.35\,\mu$m) for three different particle materials: silicon, polystyrene, and silica.}
\label{fig:simulation}
\end{figure}
%------------------------------------------------

Our theoretical findings are summarized in Fig.~\ref{fig:simulation}. For calculations of the SAM and OAM densities (Fig.~\ref{fig:simulation}(a)), we used the canonical expressions~\cite{bliokh_prl_2017}. The results agree with those in~\cite{picardi_o_2018}. Interestingly, the majority (about 96\% at the fiber surface, $d=0$) of the $z$-component of the total angular momentum comprises the spin part. Nevertheless, we expect a significant mechanical effect of the azimuthal force associated with OAM. Indeed, as shown in Fig.~\ref{fig:simulation}(b), the orbiting frequency is expected to reach about 57~Hz/W for a $1\,\mu$m (in diameter) polystyrene particle. As one can see in the inset, the maximum frequency scales with the refractive index: it equals 11~Hz/W for silicon dioxide ($n=1.45$) and 450~Hz/W for silicon ($n=3.67$). In practice, one should also consider the Brownian motion, which is inversely proportional to $R_{\rm p}$: smaller particles would exhibit longer thermal displacements and, therefore, a weaker interaction with the evanescent field, which decreases dramatically with the distance from the fiber,~$d$. As a reasonable compromise, we chose to use polystyrene beads with a diameter $2R_{\rm p}=3\,\mu$m. Under these conditions, the expected frequency for CP input is $|\tilde f_{\rm CP}|\approx21.2$~Hz/W.

%------------------------------------------------
\begin{figure}
\centering
\includegraphics[width=0.9\linewidth]{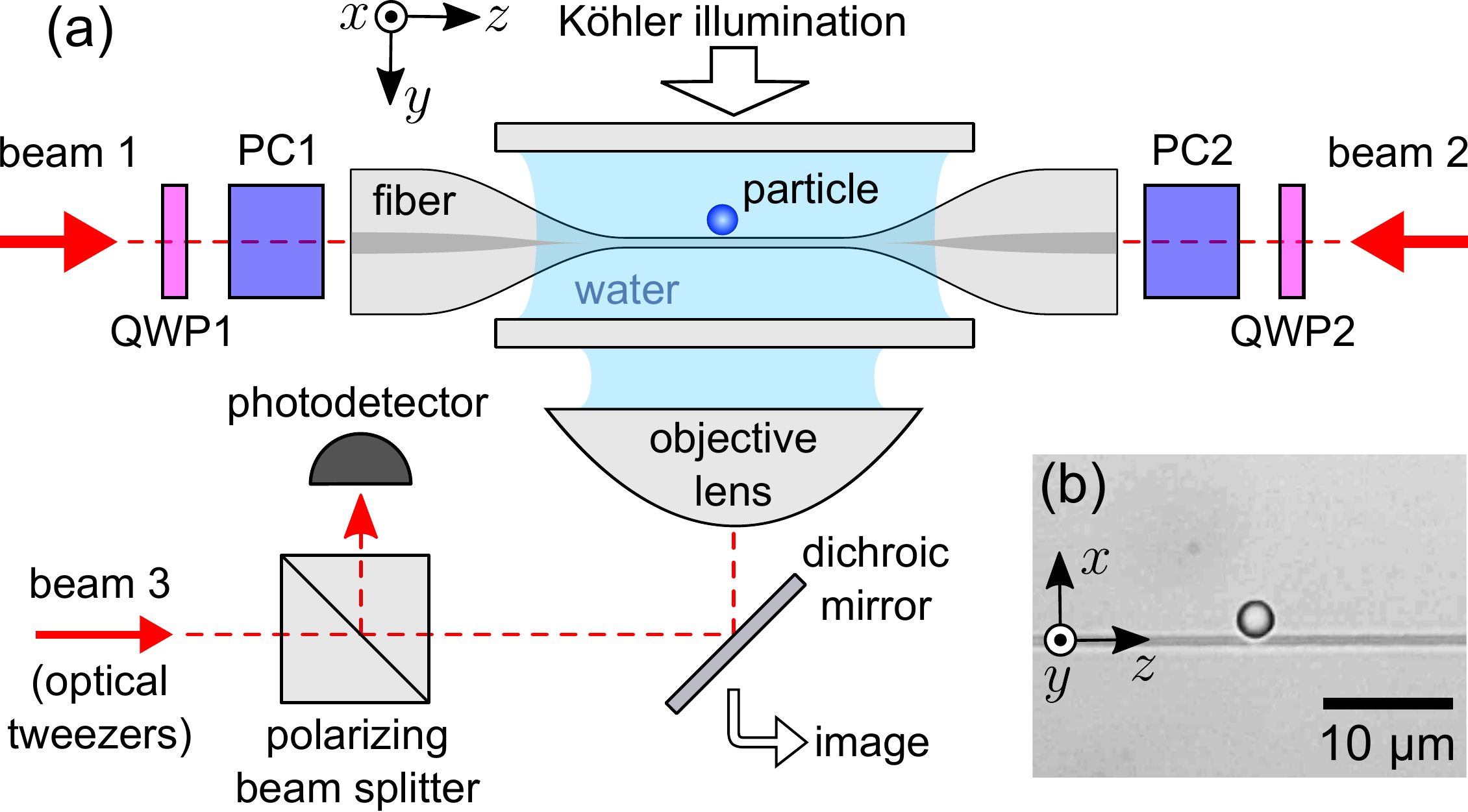}
\caption{(a)~Experimental setup: once the polarization transformations in the tapered fiber are reversed by the polarization compensators, PC1 and PC2, we set the values of $\sigma_1$ and $\sigma_2=-\sigma_1$, by rotating the quarter-wave plates, QWP1 and QWP2. (b)~Transmission image of a 3-$\mu$m polystyrene particle optically captured at the waist of a nanofiber.}
\label{fig:setup}
\end{figure}
%------------------------------------------------

Our experimental setup is sketched in Fig.~\ref{fig:setup}(a). The nanofiber is fabricated by controlled heating and pulling~\cite{ward_rsi_2014} of a step-index single-mode optical fiber (SM980G80 by Thorlabs,~Inc.). The small tapering angles of 3~mrad provide adiabatic coupling~\cite{love_ieee_1991, jung_oe_2008} between the fundamental modes in the fiber pigtails and those in the 2-mm-long cylindrical waist region having a radius of $R_{\rm f}=0.33\pm0.04\,\mu$m (measured over a set of 5 nanofibers). The fiber pigtails are coupled to non-interfering, collimated Gaussian beams~1 and~2 from a laser (Ventus, Laser Quantum Ltd.) with continuous emission at $\lambda=1.064\,\mu$m. The initial linear polarization of the beams (along~$x$ and $y$ for beams $1$ and $2$, respectively) is changed into elliptical by means of two quarter-wave plates, QWP1 and QWP2, with their slow axes oriented at equal angles, $\theta_{\rm QWP1}=\theta_{\rm QWP2}=\theta$, with respect to~$x$, measured from the point of view of the receiver. This results in $\sigma=\sin 2\theta$.

A newly made nanofiber sample is immersed into $0.3$~mL of deionized water with 3-$\mu$m polystyrene particles (Phosphorex, Inc.) and sandwiched between two parallel glass cover slips. The sample is imaged by a video camera (DCC3240C by Thorlabs, Inc.) through a water-immersion objective lens (Zeiss Plan-Apochromat, $63\times/1.00$w) under K{\"o}hler illumination, see Fig.~\ref{fig:setup}(b). Individual particles are picked up from the bottom slip using an optical tweezers realized by focusing the collimated beam~3 (from the same laser) with the same objective lens. The polarizing beam-splitter cube transmits $y$-polarized beam~3 and is subsequently used for detection (Si amplified photodetector PDA10A2 by Thorlabs,~Inc.) of the laser light escaping from the nanofiber due to scattering by the particle.

Due to uncontrolled bends, twists or geometrical inhomogeneities, the fiber does not maintain polarization of guided light. In order to control the polarization state at the nanofiber waist, we reverse the unknown polarization transformations for both beams using two free-space compensators, PC1 and PC2. The compensation procedure described elsewhere~\cite{tkachenko_arxiv_2019} is based on self-scattering from the waist imaged by a second video camera, replacing the photodetector for this purpose. 

%------------------------------------------------
\begin{figure}
\centering
\includegraphics[width=0.9\linewidth]{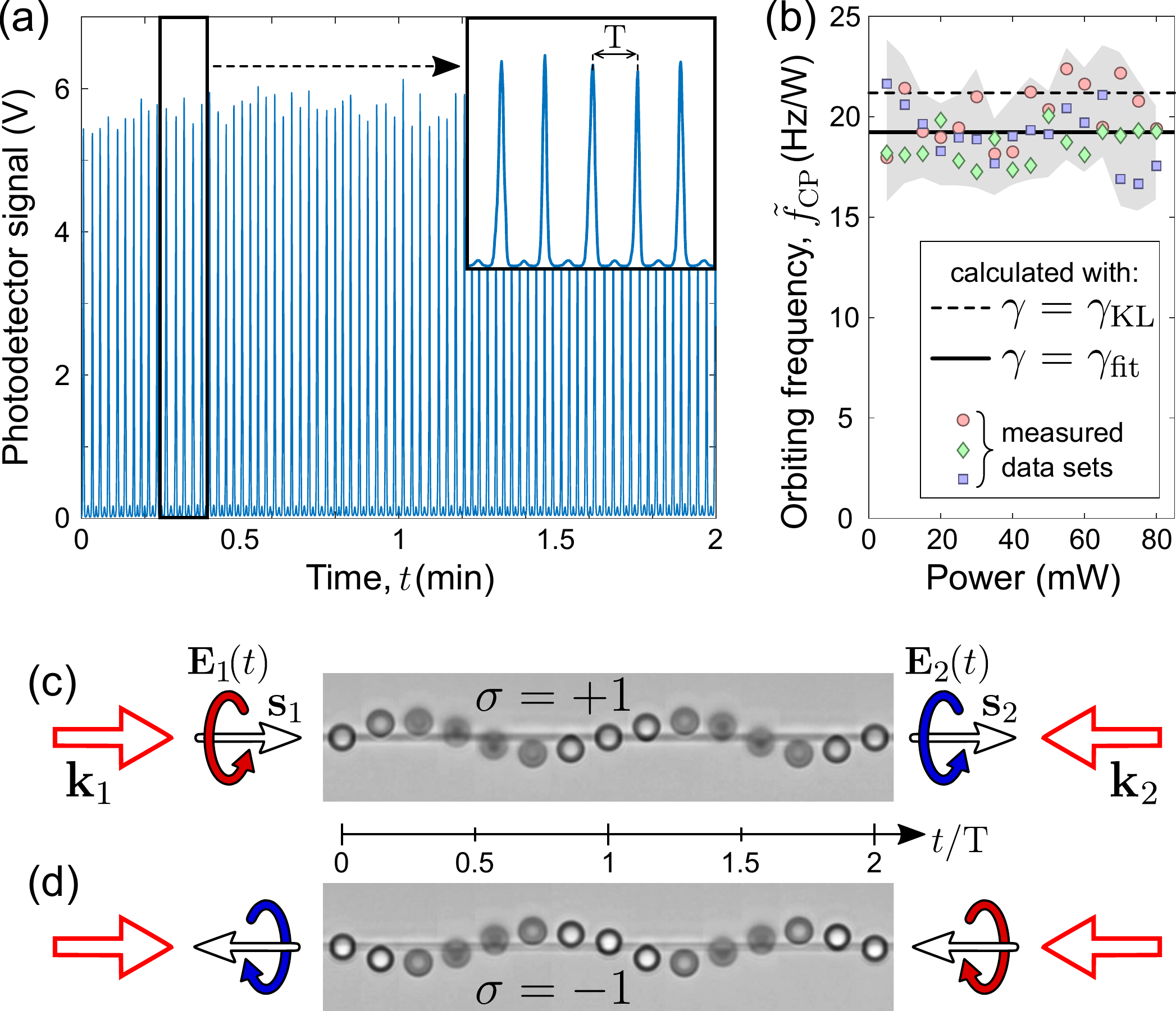}
\caption{Experimental results for a CP input (R~=~right, L~=~left). (a)~Beatings in the detector signal acquired with an optical power of 15~mW in each beam. The zoomed-in view (inset) shows the local period,~T. (b)~Orbiting frequency vs. power in each beam, at $\sigma=+1$. Markers: measured data sets for $3$~samples; gray area: combined standard deviation range. Dashed line: frequency expected for the drag coefficient $\gamma=\gamma_{\rm KL}$; solid line: the best fit to the data with $\gamma=\gamma_{\rm fit}$. (c),(d)~Time-lapse compilation of images for $\sigma=+1$~(c) and $\sigma=-1$~(d).}
\label{fig:CP}
\end{figure}
%------------------------------------------------

Experimental results with $|\sigma|=1$ are shown in Fig.~\ref{fig:CP}. Orbital motion of the particle around the fiber causes clear quasiperiodical beatings of the measured voltage, see~Fig.~\ref{fig:CP}(a). The orbiting frequency, $\tilde f_{\rm CP}$, scales linearly with optical power, as summarized in Fig.~\ref{fig:CP}(b) for three different nanofibers. The data were fitted to Eq.~\ref{eq:f_CP} with an adjustable drag coefficient, $\gamma = \gamma_{\rm fit}$. The resultant frequency, $\tilde f_{\rm CP, fit}=19.2$~Hz/W, is lower than the expected value by about 9\%, a small discrepancy given the complexity of the hydrodynamic problem, a complete solution of which is beyond the scope of this study.

When the sign of $\sigma$ is reversed, the particle rotates in the opposite direction, with nearly the same period,~T, as demonstrated by the time-lapse compilations of images in Fig.~\ref{fig:CP}(c),(d), where ${\bf s}_{1,2}$ are the $\bf k$-projections of the photon spin and the curved arrows denote the rotation of the electric field vector, ${\bf E}$, in the $xy$ plane for each beam, from the point of view of the receiver.

%------------------------------------------------
\begin{figure}
\centering
\includegraphics[width=1\linewidth]{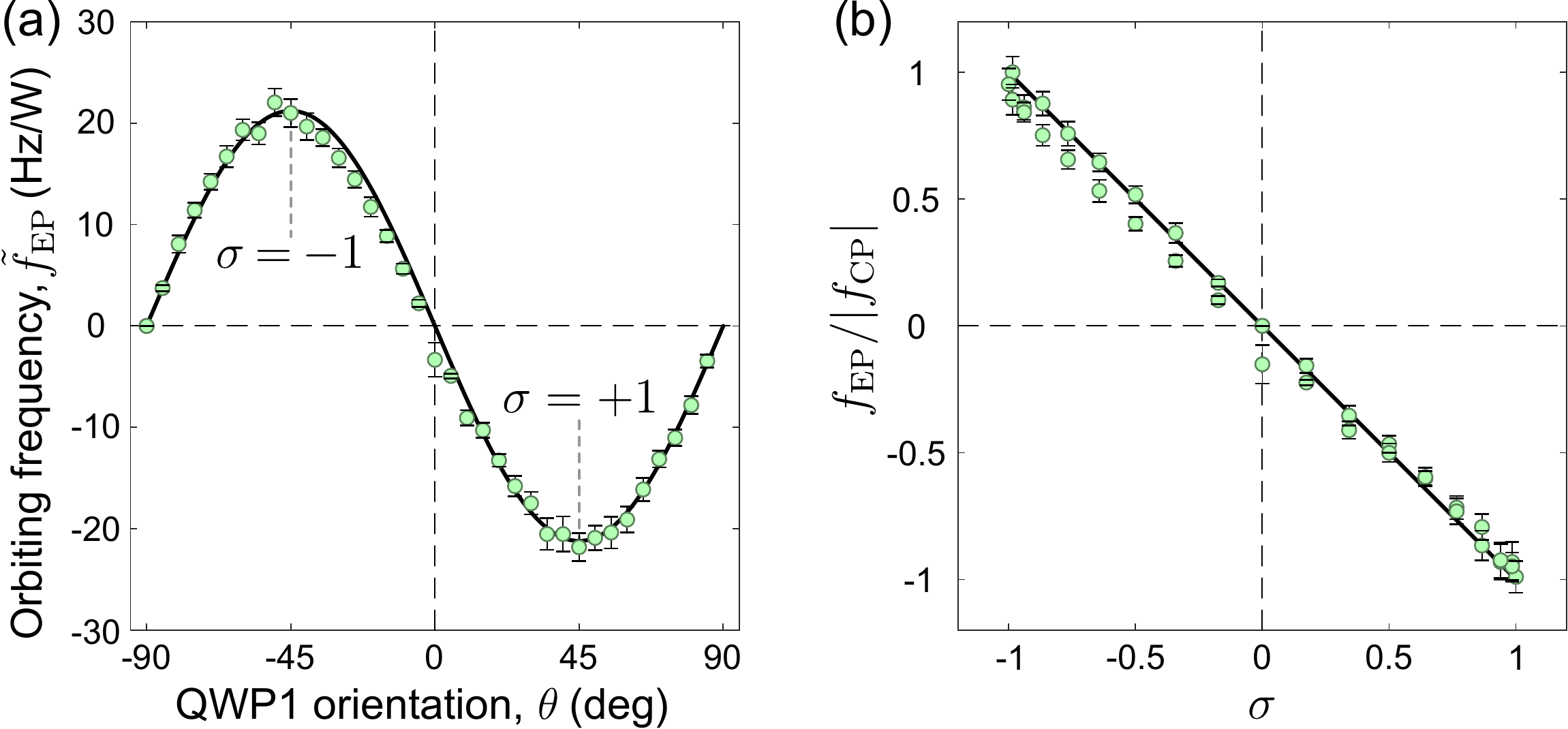}
\caption{Experimental results for an EP input and 15~mW power in each beam. (a),(b)~Markers: measured orbiting frequency versus QWP1 orientation~(a) or the $\bf k$-projection of SAM in beam~1~(b). Solid lines: simulation using Eq.~\ref{eq:f_EP}, with $\sigma=\sin2\theta$.}
\label{fig:EP}
\end{figure}
%------------------------------------------------

The results for $\sigma\neq1$ are presented in Fig.~\ref{fig:EP}, where solid lines show the simulated  frequency, $\tilde f_{\rm EP}(\sigma)$, and each error bar is the standard deviation range for at least 20~T duration. For this data set, Eqs.~{\ref{eq:f_EP},\ref{eq:f_CP},\ref{eq:gamma_KL} were applied, without adjustable parameters. As confirmed by Fig.~\ref{fig:EP}(b), the transverse spin-dependent radiation force on the particle is proportional to the SAM projection on the propagation direction, with opposite sign.
The observed light-induced rotation is antiparallel to the azimuthal component of the energy flow around the nanofiber~\cite{le_kien_pra_2013}. This counter-intuitive 'negative' radiation torque (OAM-induced) is due to the dominant forward scattering. This is associated with multipolar interference in Mie scattering from large enough particles, $R_{\rm p}>\lambda/(2\pi n_{\rm m})\approx0.13\,\mu$m.
The associated forward scattering of light relates our findings to previous demonstrations of 'negative' radiation forces~\cite{sukhov_prl_2011,chen_natPhot_2011,demore_prl_2014}.

As a curious detail, we note that $\sigma$ influences not only the frequency, but also the particle's trajectory. For CP input ($|\sigma|=1$), it is close to a circle in the $xy$ plane. When the polarization is elliptical ($|\sigma|<1$), the trajectory acquires a figure-of-eight shape, with longer trips along $z$ for smaller $|\sigma|$. This distortion is due to the lack of axial symmetry in the intensity distribution for counterpropagating elliptically polarized modes~\cite{le_kien_oc_2004}. Indeed, for $|\sigma|$ close to zero, the intensity maxima for beams~1 and~2 are aligned parallel to the $x$ and $y$ axes, respectively. Hence, the particle is accelerated towards $z>0$ or $z<0$ when passing through the $xz$ or $yz$ planes.

Here, we presented a clear experimental demonstration of a transverse, spin-dependent radiation force acting on material objects in evanescent electromagnetic fields. In contrast to previous studies on the subject, we used optical nanofibers, which provide extraordinarily clean experimental conditions, with high visibility and repeatability of measurements. An indispensable prerequisite of this experiment was the complete polarization control of light at the nanofiber waist. In addition to its use for verification of the above fundamental concept, the examined microparticle-nanofiber system could find an application in microfluidics, e.~g. as an optically addressed rotary pump.

\noindent {\bf Funding.} This work was supported by Okinawa Institute of Science and Technology Graduate University; G.~T. was supported by the Japan Society for the Promotion of Science (JSPS) as an International Research Fellow (Standard, ID~No.P18367).

\noindent {\bf Acknowledgement.} We thank J.~M.~Ward and K.~Karlsson for managing the fiber pulling rig.

\bibliographystyle{iopart-num}
\bibliography{biblio}

\providecommand{\newblock}{}
\begin{thebibliography}{10}
\expandafter\ifx\csname url\endcsname\relax
  \def\url#1{{\tt #1}}\fi
\expandafter\ifx\csname urlprefix\endcsname\relax\def\urlprefix{URL }\fi
\providecommand{\eprint}[2][]{\url{#2}}

\bibitem{friese_nature_1998}
Friese M~E~J, Nieminen T~A, Heckenberg N~R and Rubinsztein-Dunlop H 1998 {\em
  Nature\/} {\bf 394} 348--350

\bibitem{oneil_prl_2002}
{O'Neil} A~T, MacVicar I, Allen L and Padgett M~J 2002 {\em Phys. Rev. Lett.\/}
  {\bf 88} 053601

\bibitem{garceschavez_prl_2003}
{Garc{\'e}s-Ch{\'a}vez} V, McGloin D, Padgett M~J, Dultz W, Schmitzer H and
  Dholakia K 2003 {\em Phys. Rev. Lett.\/} {\bf 91} 093602

\bibitem{zhao_prl_2007}
Zhao Y, Edgar J~S, Jeffries G~D~M, McGloin D and Chiu D~T 2007 {\em Phys. Rev.
  Lett.\/} {\bf 99} 073901

\bibitem{baranova_jetp_1994}
Baranova N~B, Savchenko A~Y and Zel'dovich B~Y 1994 {\em JETP Lett.\/} {\bf 59}
  232--234

\bibitem{bliokh_prl_2008}
Bliokh K~Y, Gorodetski Y, Kleiner V and Hasman E 2008 {\em Phys. Rev. Lett.\/}
  {\bf 101} 030404

\bibitem{sukhov_natPhot_2015}
Sukhov S, Kajorndejnukul V, {Rezvani Naraghi} R and Dogariu A 2015 {\em Nat.
  Photonics\/} {\bf 9} 809--812

\bibitem{bliokh_nc_2014}
Bliokh K~Y, Bekshaev A~Y and Nori F 2014 {\em Nat. Commun.\/} {\bf 5} 3300

\bibitem{antognozzi_nPhys_2016}
Antognozzi M, Bermingham C~R, Harniman R~L, Simpson S, Senior J, Hayward R,
  Hoerber H, Dennis M~R, Bekshaev A~Y, Bliokh K~Y and Nori F 2016 {\em Nat.
  Phys.\/} {\bf 12} 731--735

\bibitem{liu_prl_2018}
Liu L, {Di Donato} A, Ginis V, Kheifets S, Amirzhan A and Capasso F 2018 {\em
  Phys. Rev. Lett.\/} {\bf 120} 223901

\bibitem{tong_oc_2012}
Tong L, Zi F, Guo X and Lou J 2012 {\em Opt. Commun.\/} {\bf 285} 4641--47

\bibitem{le_kien_pra_2006}
{Le~Kien} F, Balykin V~I and Hakuta K 2006 {\em Phys. Rev. A\/} {\bf 73} 053823

\bibitem{brambilla_ol_2007}
Brambilla G, Murugan G~S, Wilkinson J~S and Richardson D~J 2007 {\em Opt.
  Lett.\/} {\bf 32} 3041

\bibitem{xu_njp_2012}
Xu L, Li Y and Li B 2012 {\em New. J. Phys.\/} {\bf 14} 033020

\bibitem{maimaiti_sr_2015}
Maimaiti A, Truong V~G, Sergides M, Gusachenko I and {Nic Chormaic} S 2015 {\em
  Sci. Rep.\/} {\bf 5} 9077

\bibitem{frawley_oe_2014}
Frawley M~C, Gusachenko I, Truong V~G, Sergides M and {Nic Chormaic} S 2014
  {\em Opt. Express\/} {\bf 22} 16322--34

\bibitem{maimaiti_sr_2016}
Maimaiti A, Holzmann D, Truong V~G, Ritsch H and {Nic Chormaic} S 2016 {\em
  Sci. Rep.\/} {\bf 6} 30131

\bibitem{lei_prappl_2019}
Lei F, Tkachenko G, Ward J~M and {Nic Chormaic} S 2019 {\em Phys. Rev. Appl.\/}
  {\bf 11}(6) 064041

\bibitem{joos_oe_2019}
Joos M, Bramati A and Glorieux Q 2019 {\em Opt. Express\/} {\bf 27}
  18818--18830

\bibitem{tkachenko_arxiv_2019}
Tkachenko G, Lei F and {Nic Chormaic} S 2019 {\em arXiv:1907.04533\/}

\bibitem{le_kien_pra_2013}
{Le Kien} F and Rauschenbeutel A 2013 {\em Phys. Rev. A\/} {\bf 88} 063845

\bibitem{le_kien_pra_2017}
{Le~Kien} F, Busch T, Truong V~G and {Nic Chormaic} S 2017 {\em Physical Review
  A\/} {\bf 96} 023835

\bibitem{barton_jap_1988}
Barton J~P, Alexander D~R and Schaub S~A 1988 {\em J. Appl. Phys.\/} {\bf 64}
  1632--1639

\bibitem{almaas_josab_1995}
Almaas E and Brevik I 1995 {\em J. Opt. Soc. Am. B\/} {\bf 12} 2429--2438

\bibitem{le_kien_oc_2004}
{Le~Kien} F, Liang J~Q, Hakuta K and Balykin V~I 2004 {\em Opt. Commun.\/} {\bf
  242} 445--455

\bibitem{marchington_oe_2008}
Marchington R~F, Mazilu M, Kuriakose S, {Garc{\'e}s-Ch{\'a}vez} V, Reece P~J,
  Krauss T~F, Gu M and Dholakia K 2008 {\em Opt. Express\/} {\bf 16} 3712--3726

\bibitem{krishnan_pf_1995}
Krishnan G~P and Leighton D~T 1995 {\em Phys. Fluids\/} {\bf 7} 2538--2545

\bibitem{bliokh_prl_2017}
Bliokh K~Y, Bekshaev A~Y and Nori F 2017 {\em Phys. Rev. Lett.\/} {\bf 119}
  073901

\bibitem{picardi_o_2018}
Picardi M~F, Bliokh K~Y, Rodr{\'\i}guez-Fortu{\~n}o F~J, Alpeggiani F and Nori
  F 2018 {\em Optica\/} {\bf 5} 1016--1026

\bibitem{ward_rsi_2014}
Ward J~M, Maimaiti A, Le V~H and {Nic Chormaic} S 2014 {\em Rev. Sci.
  Instrum.\/} {\bf 85} 111501

\bibitem{love_ieee_1991}
Love J~D, Henry W~M, Stewart W~J, Black R~J, Lacroix S and Gonthier F 1991 {\em
  IEE Proceedings J - Optoelectronics\/} {\bf 138} 343--354

\bibitem{jung_oe_2008}
Jung Y, Brambilla G and Richardson D~J 2008 {\em Opt. Express\/} {\bf 16}
  14661--14667

\bibitem{sukhov_prl_2011}
Sukhov S and Dogariu A 2011 {\em Phys. Rev. Lett.\/} {\bf 107} 203602

\bibitem{chen_natPhot_2011}
Chen J, Ng J, Lin Z and Chan C~T 2011 {\em Nat. Photonics\/} {\bf 5} 531--534

\bibitem{demore_prl_2014}
D{\'e}mor{\'e} C~E~M, Dahl P~M, Yang Z, Glynne-Jones P, Melzer A, Cochran S,
  MacDonald M~P and Spalding G~C 2014 {\em Phys. Rev. Lett.\/} {\bf 112} 174302

\end{thebibliography}
\end{document}